\documentclass[11pt]{article}

\usepackage[main, final]{neurips_2026}

\usepackage{microtype}
\usepackage{graphicx}
\usepackage{enumitem}
\usepackage{subcaption}
\usepackage{booktabs} 
\usepackage{filecontents}

\usepackage{hyperref}

\newcommand{\R}{\mathbb{R}}
\newcommand{\OO}{\mathcal{O}}
\newcommand{\poly}{{\rm poly}}
\newcommand{\E}{\mathbb{E}}

\newcommand{\CH}{\mathsf{CH}}

\newcommand{\eps}{\varepsilon}

\usepackage{amsmath}
\usepackage{amssymb}
\usepackage{mathtools}
\usepackage{amsthm}

\usepackage[capitalize,noabbrev]{cleveref}

\theoremstyle{plain}
\newtheorem{theorem}{Theorem}[section]

\newtheorem{lemma}[theorem]{Lemma}

\theoremstyle{definition}
\newtheorem{definition}[theorem]{Definition}

\theoremstyle{fact}
\newtheorem{fact}[theorem]{Fact}
\theoremstyle{claim}

\theoremstyle{remark}
\newtheorem{remark}[theorem]{Remark}

\usepackage[textsize=tiny]{todonotes}

\title{Fast Approximate $\ell_p$ Chamfer Distance via Lopsided Embeddings and Structured JL}

\date{}

\author{%
  Ying Feng \\
  MIT \\
  \texttt{yingfeng@mit.edu} \\
  \And
  David P. Woodruf \\
  Carnegie Mellon University and Google Research \\
   \texttt{dwoodruf@andrew.cmu.edu}\\
}

\begin{document}

\maketitle

\begin{abstract}
For two $d$-dimensional point sets $A,B$  of size $n$, the Chamfer distance from $A$ to $B$ is defined as $\text{CH}(A,B)=\sum_{a \in A} \min_{b \in B} \mathsf{dist}(a, b)$, where $\mathsf{dist}$ is some
underlying distance measure.
We present efficient algorithms for approximating the Chamfer distance under the $\ell_p$ norm  for $1 < p \le 2$. For the general case $1 < p < 2$, we utilize lopsided $\ell_p$ to $\ell_1$ embeddings with weak guarantees. We show that they are sufficient for preserving the Chamfer distance. In the high-dimensional regime, we use Fast Matrix Multiplication techniques to speed up the lopsided embeddings. These lead to $\mathcal{O}(nd\cdot \poly(\log \log n, \log(1/\varepsilon))/\varepsilon^2)$ total time.   For the specific Euclidean  case ($p=2$), we leverage the Fast Johnson-Lindenstrauss Transform based on Toeplitz matrices and re-analyze the previous $\ell_1$-specific Chamfer algorithm in $\ell_2$. These achieve a runtime matching the recent state-of-the-art $\ell_1$ result of $\mathcal{O}(nd(\log \log n + \log(1/\varepsilon))/\varepsilon^2)$, improving upon the previous $\mathcal{O}(nd \log n/\varepsilon^2)$ bound for $\ell_2$. We also give additional results for the angular distance and upper and lower bounds in the streaming setting. 
\end{abstract}
\section{Introduction}

The Chamfer distance is a popular measurement for evaluating the similarity between two point sets, finding extensive application across various domains in machine learning and computer vision. Given two point sets $A, B \subset \mathbb{R}^d$ of size up to $n$, the directed Chamfer distance under the $\ell_p$ norm is defined as:

$$ \text{CH}_p(A, B) := \sum_{a \in A} \min_{b \in B} \|a - b\|_p. $$

This distance serves as a fundamental tool for quantifying dissimilarity when the exact correspondence between points is unknown or computationally intractable, making it a faster and often preferred alternative to metrics like the Earth-Mover Distance (Wasserstein Distance) \cite{Kusner2015}. 

In computer vision, the Chamfer distance is indispensable. It is widely used for tasks such as 3D shape reconstruction from single images \cite{Fan2017}, point cloud registration, and object detection. It often serves as a differentiable loss function within deep learning pipelines, enabling end-to-end training for geometric tasks \cite{athitsos2003estimating,sudderth2004visual,fan2017point,jiang2018gal}. Beyond vision, it has gained traction in natural language processing. For instance, the Relaxed Earth Mover Distance \cite{Kusner2015,atasu19a}, a variant of the Chamfer distance, is used to measure document similarity based on word embeddings. Furthermore, a closely related concept, the ``sum of maximum similarities'', where the minimum distance is replaced by the maximum inner product, underpins the efficient ColBERT passage retrieval system \cite{khattab2020colbert}.

The necessity for efficient computation is particularly acute in modern high-dimensional settings. Whether comparing high-dimensional feature embeddings generated by deep neural networks or analyzing large-scale 3D models, the naive $\mathcal{O}(n^2 d)$ computation time quickly becomes a bottleneck. This necessitates the development of fast approximation algorithms.

The first improvement over the naive  $\OO(dn^2)$-time algorithm was obtained in~\cite{sudderth2004visual}. They use the fact that $\CH(A,B)$ can be computed by performing $|A|$ nearest neighbor queries in a data structure storing $B$. Using this framework and the state of the art approximate nearest neighbor algorithms designed for low-dimensional regime, it leads to an $(1+\epsilon)$-approximate algorithm with $\OO\left(n (1/\eps)^{\OO(d)} \log n\right)$ running time~\cite{AM98}, or a similar bound that scales exponentially in the dimension (e.g~\cite{elkin23a}), and $\OO \left(dn^{1+\frac{1}{2 (1+\epsilon)^2 -1}}\right)$ running time in high dimensions~\cite{andoni2015optimal}. However,  the first bound suffers from exponential
dependence on the dimension, while the second bound is significantly subquadratic only for large approximation factors.
 Recent advancements have made significant strides in accelerating the computation. \cite{BIJ24} provided the first near-linear time $(1 + \varepsilon)$-approximation in $\mathcal{O}(nd \log(n)/\varepsilon^2)$ time for $\ell_1$ and $\ell_2$. Subsequently, \cite{Feng2025} introduced a highly efficient $(1 + \varepsilon)$-approximation algorithm specifically for the $\ell_1$ Chamfer distance, achieving a runtime of $\mathcal{O}(nd(\log \log n + \log(1/\varepsilon))/\varepsilon^2)$. However, extending this efficiency gain to the $\ell_2$ norm and the broader range of $1 < p \le 2$ remained an open challenge. This is significant because the choice of $p$ significantly impacts the properties of the distance measure and its suitability for specific tasks, and many critical applications rely on the Euclidean ($\ell_2$) norm or other $\ell_p$ norms ($1 < p \le 2$). 

\subsection{The Importance of the intermediate range $1<p\leq2$.}
\paragraph{Euclidean ($\ell_2$) Norm:} The $\ell_2$ norm is the standard geometric distance and is fundamental in applications involving physical space, such as robotics and 3D modeling. It is also the default metric in many embedding spaces (e.g., $\ell_2$-normalized embeddings in contrastive learning). Also, it is closely related to the MaxSim similarity used in the ColBERT system 

\paragraph{$\ell_p$ Norm for Intermediate $p \in (1, 2)$:} While $\ell_1$ (Manhattan distance) promotes sparsity and is robust to outliers, and $\ell_2$ is sensitive to large deviations, intermediate $\ell_p$ norms ($1 < p < 2$) offer a crucial trade-off. They provide a tunable balance between the robustness of $\ell_1$ and the rotational invariance of $\ell_2$. This provides a natural theoretical motivation for studying
intermediate $\ell_p$ norms, as different choices of $p$
may be appropriate for different notions of dissimilarity.

To provide a small empirical illustration of how the choice
of $p$ can affect the ordering induced by Chamfer distance,
we compare $\mathrm{CH}_p$ for $p=1,1.5,2$ against exact
Euclidean Earth Mover Distance (EMD) on synthetic point-cloud
pairs.

Let $\CH_p$ denote the Chamfer distance under the $\ell_p$ norm. We empirically show that $\CH_p$ for $p=1.5$ serves as a better surrogate for EMD than $p=1, 2$ on a family of synthetic, naturally motivated point-cloud pairs. In this experiment, we generate point-clouds pairs $(A, B)$ in $\R^{d=64}$, each containing $n=40$ points. For each pair, we first sample a reference cloud $B$ i.i.d. from the standard Gaussian $\mathcal{N}(0, I)$. We then generate $A$ from $B$ using one of the following three stochastic corruption approaches. In all approaches, we use a parameter $t\in[0,1]$ to controls the corruption strength:

\begin{enumerate}[leftmargin=2em]
    \item \textbf{Dense-shift.} We first draw a random unit vector $u$ and define a shift magnitude $m=m(t)$.
Then every point $a_i \in A$ is shifted from $b_i$ by an offset vector $(m+ \alpha_i)u + \beta_i$, where $\alpha_i, \beta_i$ are Gaussian noise sampled i.i.d. for each $i$. 

\item  \textbf{Many-to-one-collapse.} We start from $A$ being a global shift of $B$ by the same random offset vector. Then we choose a random fraction $f=f(t)+\alpha$ of points in $A$ to collapse, and a random fraction $g=g(t)+\beta$ of points in $B$ to collapse to, where $\alpha, \beta$ are again Gaussian noise, and $f \gg g$. For each chosen point in $A$, it is moved to one of the chosen points in $B$ uniformly at random. This creates a many-to-one collapse of a subset of points toward a small set of hubs, plus a small common drift.

\item \textbf{Sparse-spikes.} We start from $A=B$. Then we choose a small random fraction $a=a(t)+\alpha$ of points in $A$ to corrupt, and for each chosen point, we choose a small random subset of coordinates, and perturb it by a Gaussian offset whose variance is $b = b(t)+\beta$, where $\alpha, \beta$ are again Gaussian noise. Thus only a sparse subset of points is perturbed.

\end{enumerate}

These three types of corruptions are intended to capture, respectively, diffuse geometric mismatch, loss of correspondence structure, and sparse outlier-like corruption. To make the three corruption types similarly representative, we do not directly use the same parameter $t$ across families. Instead, we compute a calibration curve for $t \mapsto \text{Empirical Expectation of EMD}$. We then choose $10$ target EMD levels, equally spaced in the interval $[0.8,2.6]$, and for each family invert the calibration curve to obtain the value of $t$ that approximately matches that target EMD level.

For each random seed, we use: $10$ target EMD levels, $3$ corruption families, and $15$ independent samples -- a total of $450$ point-cloud pairs per seed. For each generated pairs $(A, B)$ , we compute EMD and a symmetric version of $\CH_p$: $\CH_p(A, B) + \CH_p(B, A)$, for $p=1, 1.5, 2$. This induces a ranking among $450$ pairs for each measurement.  
To evaluate how well $\CH_p$ serves as a surrogate for  EMD, we compute the Spearman rank correlation between the ranking induced by $\CH_p$ and the  ranking induced by EMD. A larger Spearman correlation means that the corresponding Chamfer variant better preserves the ordering induced by exact EMD, which is the relevant notion of surrogate quality here. Our result is shown in \cref{tab:spearman}.
\begin{table*}[t]
\centering
\caption{Spearman correlation between symmetric $\ell_p$-Chamfer distance and exact Euclidean EMD across 5 random seeds. Each seed uses $n=40$, $d=64$, and $450$ generated point-cloud pairs. In every seed, the intermediate choice $p=1.5$ attains the highest correlation.}
\label{tab:spearman}
\resizebox{0.72\textwidth}{!}{%
\begin{tabular}{@{}cccc@{}}
\toprule
\textbf{Seed} & \textbf{Average Spearman ($p=1.0$)} & \textbf{Average Spearman ($p=1.5$)} & \textbf{Average Spearman ($p=2.0$)} \\
\midrule
0 & 0.8783 & \textbf{0.8817} & 0.8802 \\
1 & 0.8895 & \textbf{0.8916} & 0.8870 \\
2 & 0.8898 & \textbf{0.8917} & 0.8889 \\
3 & 0.8894 & \textbf{0.8903} & 0.8896 \\
4 & 0.8761 & \textbf{0.8786} & 0.8749 \\
\bottomrule
\end{tabular}%
}
\end{table*}

We observe that $p=1.5$ achieves the highest Spearman
correlation in each of the five random seeds, i.e., it achieves the strongest agreement with EMD. This is exactly the type of effect one should expect from an intermediate norm. The $\ell_1$-Chamfer distance is relatively more influenced by diffuse high-dimensional shifts, while $\ell_2$-Chamfer is relatively more affected by sparse large perturbations and mismatch structure created by collapse. The intermediate choice $p=1.5$ provides a compromise between these two behaviors.

More broadly, Chamfer distance is useful beyond serving as
a surrogate for EMD, since its distinct geometry may itself
be desirable. For instance, \cite{Fan2017} (Fig. 3. c, d) illustrates that,
in settings with categorical uncertainty, the mean shape
under a squared-Euclidean Chamfer loss can preserve the
possible locations of uncertain structures, whereas the
EMD mean can be substantially distorted.
This demonstrates that Chamfer can encode a preferable notion of point-set dissimilarity; understanding how it varies with the underlying $\ell_p$
 metric is therefore of independent interest.

\subsection{Our Contributions}
In this work, we present fast algorithms for approximating the Chamfer distance for the general case of $1 < p \le 2$. We also empirically validate our results by experiments on both synthetic and real-world data. 
We leverage embedding and dimensionality reduction techniques tailored to different $\ell_p$ norms. Our results are summarized in Table \ref{tab1}.

\begin{table*}[t]
\centering
\caption{Comparison of $(1+\varepsilon)$-approximate Chamfer Distance Algorithms for $n$ points in $d$ dimensions. Our work is the first to achieve near-linear time for the intermediate range $1<p\leq2$, extend the $\ell_1$ state-of-the-art runtime to $\ell_2$ and ColBERT MaxSim, and support space-efficient streaming. The time complexity of intermediate norms and MaxSim assumes $d \geq \poly\log(n)$.}
\label{tab1}
\resizebox{\textwidth}{!}{%
\begin{tabular}{@{}llccc@{}}
\toprule
\textbf{Algorithm / Reference} & \textbf{Supported Metric} & \textbf{Time Complexity} & \textbf{Space Complexity} & \textbf{Dynamic Streaming} \\
\midrule
Naive Exact & Any $\ell_p$ ($p \ge 1$) & $\mathcal{O}(n^2 d)$ & $\mathcal{O}(nd)$ & $\times$ \\
\cite{BIJ24} & $\ell_1, \ell_2$ & $\mathcal{O}\left(nd \log(n) / \varepsilon^2\right)$ & $\mathcal{O}(nd)$ & $\times$ \\
\cite{Feng2025} & $\ell_1$ & $\mathcal{O}\left(nd(\log \log n + \log(1/\varepsilon))/\varepsilon^2\right)$ & $\mathcal{O}(nd)$ & $\times$ \\
\midrule
\textbf{Ours (Theorem 3.4)} & \textbf{General $\ell_p$ ($1 < p < 2$)} & $\mathcal{O}\left(nd \cdot \text{poly}(\log \log n, \log(1/\varepsilon))/\varepsilon^2\right)$ & $\mathcal{O}(n \cdot \text{poly}(\log n))$ & \checkmark \\
\textbf{Ours (Theorems 4.5 \& 5.2)} & \textbf{$\ell_2$ and MaxSim} & $\mathcal{O}\left(nd(\log \log n + \log(1/\varepsilon))/\varepsilon^2\right)$ & $\mathcal{O}(n \log^2 n)$ & \checkmark \\
\midrule
\textbf{Our Lower Bound (Theorem 5.3)} & \textbf{Any $\ell_p$ ($p \ge 1$)} & \textbf{---} & \textbf{$\Omega(n)$ bits} & \textbf{---} \\
\bottomrule
\end{tabular}%
}
\end{table*}

\textbf{General Case ($1 < p < 2$).} For the general case, we utilize lopsided embeddings from $\ell_p^d$ to $\ell_1^m$ for $m = \poly(\log n, \varepsilon^{-1})$ \cite{Li2023}. These embeddings guarantee non-contraction for all $\mathcal{O}(n^2)$ pairwise distances between points in sets $A$ and $B$, while preserving nearest neighbor distances with bounded expected error.  Focusing on the high-dimensional case where $d \ge \poly(\log n)$, we show that Fast Matrix Multiplication (FMM) techniques \cite{Swartworth2026} can apply the dense embedding matrix efficiently, resulting in a runtime near-linear in the input dimension $d$. These lead to $\mathcal{O}(nd\cdot \poly(\log \log n, \log(1/\varepsilon))/\varepsilon^2)$ time algorithm for estimating the Chamfer distance for $p \in (1, 2)$. We remark that this approach also works for $p=2$, but by opening   the black-box of the algorithm in \cite{Feng2025} and combining it with appropriate dimension reduction, we can achieve an even faster runtime  for $p=2$.

\textbf{Euclidean Case ($p = 2$).} For the $\ell_2$ norm, we give an algorithm that matches the best known $\ell_1$ runtime, solving an open problem in \cite{Feng2025}.  To this end, we use a Fast Johnson-Lindenstrauss Transform (FJLT) based on structured Toeplitz matrices \cite{Krahmer2011, Hinrichs2011, Vybiral2011} to reduce the dimension from $d$ to $m = \mathcal{O}( \log^2 n/\varepsilon^{2})$, and then re-analyze the algorithmic framework of \cite{Feng2025} in the reduced dimension in the $\ell_2$ space. Although the algorithm in \cite{Feng2025} is designed to approximate the Chamfer distance in the $\ell_1$ norm, we show that it actually gives a good approximation in the reduced-dimensional $\ell_2$ space. Related ideas of analyzing tree embeddings for $\ell_2$ norm appeared in \cite{lahn25a}. However, their tree construction takes quadratic time, which is in contrast to our near-linear-time algorithm framework.  Combining this with the $\mathcal{O}(d \log m)$-time Toeplitz matrix multiplication, we achieve a total runtime of $\mathcal{O}(nd(\log \log n + \log(1/\varepsilon))/\varepsilon^2)$, matching the $\ell_1$ result and improving the previous $\ell_2$ bound.

\textbf{Extensions and Theoretical Limits.} Beyond geometric $\ell_p$ norms, we show that our techniques also apply to similarity measures used in modern LLM retrieval.
In particular, we consider the ColBERT MaxSim score, defined for token embeddings on the unit sphere as
\[
\mathrm{MaxSim}(A,B)=\sum_{a\in A}\max_{b\in B}\langle a,b\rangle .
\]
Using the identity $\|a-b\|_2^2 = 2-2\langle a,b\rangle$ for normalized vectors, we show that our Fast $\ell_2$ Chamfer
algorithm (Theorem \ref{thm:l2_chamfer}) yields a near-linear-time \emph{additive} $\OO(\varepsilon)$-approximation to $\mathrm{MaxSim}(A,B)$.

We also study the Chamfer distance in the \emph{dynamic turnstile streaming} model, where point coordinates receive additive
updates. Since both of our dimensionality reduction maps (Toeplitz FJLT and lopsided embeddings) are linear, they can be
maintained online under updates, allowing the algorithm to store only the low-dimensional sketches and run the downstream
solver at the end of the stream, reducing space from $\OO(nd)$ to $\OO(n\,\mathrm{poly}(\log n))$.
Complementing this upper bound, we prove an unconditional $\Omega(n)$ space lower bound for any randomized constant-pass
streaming algorithm achieving a constant-factor approximation (for any $p\ge 1$), via a reduction from two-party set
disjointness. We note that while a dynamic version of the Chamfer distance problem has been recently studied in \cite{G25}, they assume that the sets undergo insertions and deletions of {\it points}, while our setting allows arbitrary updates to the coordinates of points. 

\section{Preliminaries}

\subsection{Fast Sketch Application via FMM}
We need to efficiently apply dense sketch matrices used in lopsided embeddings. This is done using the Fast Matrix Multiplication (FMM) technique from \cite{Swartworth2026}.

\begin{lemma}[Fast Sketch Application, Adapted from Theorem C.1 and C.3 of \cite{Swartworth2026}] \label{lem:FMM}
Let $T \in \mathbb{R}^{m \times d}$ be a dense sketch matrix. There exists a constant $K \approx 11.76$ (related to the exponents of rectangular matrix multiplication) such that if $d \ge m^{2K}$, the product $Tv$ for any vector $v \in \mathbb{R}^d$ can be computed in time $\mathcal{O}(d \cdot \poly \log(m))$.
\end{lemma}

\subsection{Fast Johnson-Lindenstrauss Transform (FJLT)}
For the Euclidean case, we use FJLT variants over matrices in $\mathbb{R}^{m \times d}$  that provide the guarantees of the JL lemma \cite{Johnson1984}.  We highlight that we require the FJLT runtime to be $\mathcal{O}(d \log m)$, which is achieved by Toeplitz matrices \cite{Krahmer2011, Hinrichs2011, Vybiral2011},  instead of the more standard $\mathcal{O}(d \log d)$ bound often associated with full-dimensional FFTs. This is crucial to obtain the desired $\poly\log\log(n)$ overhead in the final runtime.

\begin{lemma}[Fast Johnson-Lindenstrauss Transform, \cite{Krahmer2011, Hinrichs2011, Vybiral2011}]\label{lem:reduction} Let $V \subset \mathbb{R}^d$ be a set of $N$ points. Let $\varepsilon, \delta \in (0, 1/2)$. There exists a distribution over structured matrices $T \in \mathbb{R}^{m \times d}$ where $m = \mathcal{O}( \log^2(n/\delta) / \varepsilon^2)$, such that with probability $1 - \delta$, for all $u, v \in V$:
\[ (1 - \varepsilon) \|u - v\|_2 \le \|T(u - v)\|_2 \le (1 + \varepsilon)\|u - v\|_2. \]

Moreover, $Tv$ can be computed in $\mathcal{O}(d \log m)$ time.
\end{lemma}

\subsection{The Framework of \cite{Feng2025}}
The algorithm in \cite{Feng2025} consists of three main
components: QuadTrees, Tournaments, and Rejection Sampling. 
In particular, the QuadTrees compute crude estimates $\mathcal{D}_a$ of $\mathsf{opt}_a := \min_{b \in B} \|a-b\|$ for all $a \in A$. And the efficiency of the overall algorithm depends on the approximation factor achieved by the
QuadTree step. If QuadTree provides an $F$-approximation in expectation (i.e., $\mathbb{E}[\mathcal{D}_a] \leq F\cdot \mathsf{opt}_a$),
then the overall runtime of the algorithm is dominated by $\mathcal{O}(nd\log\log n + n(d+F/\varepsilon^2)\log (F/\varepsilon^2))$.

For the $\ell_1$ metric, the authors prove $F = \mathcal{O}(\min(d, \log n ))$. This leads to  $\mathcal{O}(nd\log\log n/\varepsilon^2)$ total runtime. In Section \ref{sec:l2}, we will re-analyze the approximation factor $F$ when the framework is applied to the $\ell_2$ metric.

We also note that given the existence of $p$-stable distributions, the rest of the algorithm in \cite{Feng2025} works the same regardless of the underlying $p$-norm. These can be summarized into the following theorem:

\begin{theorem}[Adapted from \cite{Feng2025}]\label{thm:adapted}
Given a QuadTree procedure that achieves an expected approximation factor $F$, there exists an algorithm that, with .9 probability, computes a $(1 + \varepsilon)$-approximation to the Chamfer distance in total time: $\mathcal{O}(nd\log\log n + n(d+F/\varepsilon^2)\log (F/\varepsilon^2))$.
\end{theorem}

\section{The General Case ($1 < p < 2$)}

For $1 < p < 2$, we use lopsided embeddings into $\ell_1$. Unlike the $\ell_2$ case where the JL Lemma guarantees two-sided  strong concentration, the lopsided embeddings are weaker, providing only strong non-contraction  (but not dilation) and  bounded expected error. However, we show that this suffices for our need, due to the fact that the Chamfer distance is defined as the sum of the nearest neighbor distance. 

\subsection{Lopsided $\ell_p$ to $\ell_1$ Embedding for Finite Sets}

We utilize the results based on $p$-stable distributions from \cite{Li2023}.

\begin{lemma}[Lopsided Embedding for Finite Sets, adapted from \cite{Li2023}] \label{lemma:lopsided} Suppose $p \in (1, 2)$ is constant. Let $\varepsilon, \delta \in (0, 1/2)$ and $V \subset \mathbb{R}^d$ with $|V| = n$. There exists a distribution over matrices $T \in \mathbb{R}^{m \times d}$, where $m = \poly(\log n, \varepsilon^{-1}, \delta^{-1})$, satisfying:
\begin{enumerate}
    \item \textbf{(Expected Error Bound):} For any fixed $v \in \mathbb{R}^d$, $\mathbb{E}[|\|Tv\|_1 - \|v\|_p|] \le \varepsilon\|v\|_p$.
    \item \textbf{(Contraction on V):} With probability $1 - \delta$, for all $v \in V$, $\|Tv\|_1 \ge (1 - \varepsilon)\|v\|_p$.
\end{enumerate}
\end{lemma}

This is implied by the proof to Lemma 18 of \cite{Li2023}. We note that the original Lemma 18  in \cite{Li2023} is stated for the target space $\ell_q$ with $q \in (1, 2)$, whereas here we need the lemma to hold for $\ell_1$, i.e. the endpoint $q = 1$. Thus, we cannot use the prior theorem entirely as a black box:  we revisit the proof and verify that the argument extends to the target needed here. We delay the proof to Appendix \ref{sec:appendix} due to the page limit.

\subsection{The Algorithm}
The algorithm is via a reduction from the $\ell_p^d$ instance to the $\ell_1^m$ instance. We show that the weak guarantees provided by the lopsided embeddings imply good preservation of the Chamfer distance. We need to ensure non-contraction for all $n^2$ pairwise differences. Consequently, we will set $m$ in the embedding (Lemma \ref{lemma:lopsided}) to be polylogarithmic in $n$. Note that the embedding matrix is dense, thus to speed up the embedding time, we will use  the FMM technique given in Lemma \ref{lem:FMM}.

\begin{theorem} \label{thm:general_case}
Let $1 < p < 2$ be constant, $A, B \subset \mathbb{R}^d$ of size $n$, and $\varepsilon \in (0, 1/2)$. For some $m = \poly(\log n, \varepsilon^{-1})$, if $d \ge m$, then there is an algorithm that computes a $(1 + \varepsilon)$-approximation to $\text{CH}_p(A, B)$ in total time:
$$ \mathcal{O}\left(nd \cdot \poly \log(m) + T_1(n, m, \varepsilon) \right), $$
with $.9$ probability, where $T_1(n, m, \varepsilon)$ is the time to 
$(1 + \varepsilon)$-approximation the $\ell_1$ Chamfer distance in $\mathbb{R}^m$ with high probability.
\end{theorem}

\begin{proof}
We sample a lopsided embedding matrix $T$ from Lemma \ref{lemma:lopsided}. The algorithm proceeds by (1) applying the embedding matrix $T$ from $\ell_p^d$ to $\ell_1^m$ using FMM, and (2) running the $\ell_1$ Chamfer algorithm on the embedded points. 

\textbf{Runtime:} The runtime of step (2) is clearly $T_1(n, m, \varepsilon)$. We thus only need to argue that step (1) runs in time $\mathcal{O}(nd \cdot \poly \log(m))$, which follows immediately from Lemma \ref{lem:FMM}.

\textbf{Correctness:} It suffices to show that
$\text{CH}_1(TA, TB) = (1\pm \mathcal{O}(\varepsilon)) \text{CH}_p(A, B) .$
We apply Lemma \ref{lemma:lopsided} on set $V := \{a - b \mid a \in A, b \in B\}$, which has size $\mathcal{O}(n^2)$.  We show the lower and upper bound separately:

\textbf{(Lower Bound).}  By the second guarantee of $T$ in Lemma \ref{lemma:lopsided}, we can argue that with $.99$ chance,
\begin{align*}
    \text{CH}_1(TA, TB) &= \sum_{a \in A} \min_{b \in B} \|T(a-b)\|_1 \\
    &\geq (1- \varepsilon)\sum_{a \in A} \min_{b \in B} \|a-b\|_p \\
    &=  (1- \varepsilon)\text{CH}_p(A, B) 
\end{align*}
where the second step follows from the fact that for any minimizer $b^*$ of $\min_{b \in B} \|T(a-b)\|_1 $, it must hold that 
\[\|T(a-b^*)\|_1 \geq (1- \varepsilon)\|a-b^*\|_p \geq  (1- \varepsilon)\min_{b \in B} \|a-b\|_p .\] 

\textbf{(Upper Bound).} We now show that the embedded distance does not blow up. We use $b_a^*$ to denote $\arg\min_{b\in B} \| a-b\|_p$ (when there are multiple minimizers, we fix an arbitrary one of them). Using the first guarantee of Lemma \ref{lemma:lopsided} and the linearity of expectation:

\begin{align*}
&\mathbb{E} \Bigl[ \Bigl|\sum_{a \in A} \| T (a - b_a^*) \|_1  - \text{CH}_p(A, B)\Bigr|\Bigr] \\
     = &\mathbb{E} \Bigl[ \Bigl|\sum_{a \in A} \| T (a - b_a^*) \|_1  - \sum_{a \in A} \|a - b_a^*\|_p \Bigr|\Bigr] \\
    \leq &\sum_{a \in A} \mathbb{E} \Bigl[ \Bigl| \| T (a - b_a^*) \|_1  -  \|a - b_a^*\|_p \Bigr|\Bigr]  \\
    \leq &\varepsilon \cdot \sum_{a \in A} \|a - b_a^*\|_p \\
    = & \varepsilon \cdot \text{CH}_p(A, B)\\
\end{align*}
This bounds the expectation of the approximation error. By Markov's inequality, the probability of this error term exceeding $100\eps$ is only $.01$. Therefore, with probability at least $.99$, the complementary event holds, i.e.
\[|\sum_{a \in A} \| T (a - b_a^*) \|_1  - \text{CH}_p(A, B)| \leq  100\varepsilon\cdot \text{CH}_p(A, B) = \mathcal{O}(\varepsilon)\cdot \text{CH}_p(A, B).\]

On the other hand, since 
\begin{align*}
    \text{CH}_1(TA, TB) &=  \sum_{a \in A} \min_{b \in B}\| T (a - b) \|_1 \\
    &\leq \sum_{a \in A} \| T (a - b_a^*) \|_1,
\end{align*}
it follows that $\text{CH}_1(TA, TB) \leq  \text{CH}_p(A, B) + \mathcal{O}(\varepsilon)\cdot \text{CH}_p(A, B)$, which concludes the proof.

\end{proof}

\begin{remark}
    We remark that using a standard median trick, the $.9$ success probability in Theorem \ref{thm:general_case} can be boosted to $1-\delta$ for any $\delta \in (0, 1)$, by repeating the algorithm for $\log(1/\delta)$ time. 
\end{remark}

Combining this with the known Chamfer algorithm under the $\ell_1$ norm, we get the following result:

\begin{theorem}
    Let $1 < p < 2$ be constant, $A, B \subset \mathbb{R}^d$ of size $n$, and $\varepsilon \in (0, 1/2)$. For some $m = \poly(\log n, \varepsilon^{-1})$, if $d \ge m$, then there is an algorithm that computes a $(1 + \varepsilon)$-approximation to $\text{CH}_p(A, B)$ in total time $\mathcal{O}\left(nd \cdot \poly ( \log\log n + \log (1/\varepsilon)) /\varepsilon^2\right) $ with $.9$ probability.
\end{theorem}

\section{The Euclidean Case ($p = 2$)}\label{sec:l2}

We now analyze the Euclidean case by combining the FJLT with a modified analysis  of \cite{Feng2025}. As outlined in the introduction, the strategy is to first reduce the dimension using a Toeplitz structure, and then analyze the performance of the QuadTree estimator in the reduced dimension.

In this section, we let $m$ denote the target dimension obtained after the FJLT projection, where the point sets now reside. We apply Lemma \ref{lem:reduction} with the failure probability $\delta$ set to $.01$ and approximation factor set to $1$. Thus we can set $m = \mathcal{O}( \log^2n)$.  We stress that we only need a constant approximation here because the FJLT is not applied as a black-box reduction of the entire Chamfer instance; it is used only in the QuadTree stage of the \cite{Feng2025} framework, where we need a crude, expected  approximation to each nearest-neighbor distance. In other words, the constant JL distortion can be absorbed into the already-crude approximation factor.

\subsection{QuadTree Analysis in $\ell_2$}\label{sec:quadtree}
We first analyze the approximation factor $F_{\ell_2}$ achieved by the QuadTree procedure of \cite{Feng2025} when applied directly to the $\ell_2$ metric in dimension $m$. A quadtree of depth $t$   is defined by a random offset vector $z \sim [0, 2^{t}]^m$. It assigns every point in $\R^m$ with a sequence of $t$ hash values, using $t$ nested grids shifted by $z$. Concretely, for any $a \in \R^m$ and any integer $k$ such that $0 \leq k \leq t$,  it hashes 
\[h_k(a) := (\lceil\frac{a_1+z_1}{2^k}\rceil, \lceil\frac{a_2+z_2}{2^k}\rceil,  \cdots, \lceil\frac{a_m+z_m}{2^k}\rceil).\] 

\cite{Feng2025} used two independent quadtrees to estimate the nearest neighbor distance for all $a \in A$ in $\ell_1$, i.e. estimating $\min_{b\in A}\|a-b\|_1$ for all $a \in A$. We follow a similar proof structure and analyzes the behavior of quadtrees in $\ell_2$. Due to the page limit, we delay the proofs to the appendix \cref{sec:appendix-quadtree}.

\begin{lemma}\label{lem:1}

For all $a, b \in \R^m$ and $0 \leq k \leq t$, if $h_k(a) = h_k(b)$, then $\|a -b\|_2 \leq 2^k \sqrt{m}$.
\end{lemma}

\begin{lemma}\label{lem:2}
    With probability $1 - \OO(1/n)$, the following holds
simultaneously for all $a,b, k$: If $h_k(a) = h_k(b)$, then $\|a-b\|_2 \leq 2^k \cdot 3\log n$.
\end{lemma}

With the above lemmas, we can show that in expectation, two quadtrees output good estimation to $\mathsf{opt}_a := \min_{b\in A}\|a-b\|_2$ for all $a \in A$. The estimator $\mathcal{D}_a$ is defined as: 
\begin{enumerate}
    \item Identifying the smallest $\tilde{k}$ such that $h_{\tilde{k}}(a) = h_{\tilde{k}}(b)$ for some $b \in B$ across the two quadtrees.
    \item Assigning $\mathcal{D}_a := \|a-b\|_2$ for arbitrary $b$ such that $h_{\tilde{k}}(a) = h_{\tilde{k}}(b)$.
\end{enumerate}

\begin{lemma}\label{lem:4}
    With probability $1 - \OO(1/n)$, it holds for all $a \in A$ that $\E{\mathcal{D}_a} \leq F \cdot \mathsf{opt}(a)$, where $F  = \OO(\min(m,\sqrt{m}\log n))$.
\end{lemma}

    
    

\subsection{Algorithm for $\ell_2$ Chamfer Distance}

As mentioned in the introduction, it can be verified that when shifting from the $\ell_1$ norm to $\ell_2$,  the rest of the algorithm in \cite{Feng2025} works the same. Thus all it remains is the combine the FJLT with the analysis above.

\begin{theorem}[Fast $\ell_2$ Chamfer Distance]\label{thm:l2_chamfer} Given $A, B \subset \mathbb{R}^d$ of size $n$, and $\varepsilon \in (0, 1/2)$. There is an algorithm that computes a $(1 + \varepsilon)$-approximation to $\text{CH}_2(A, B)$ with $.9$ probability in time $\mathcal{O}\left(nd(\log \log n + \log(1/\varepsilon))/\varepsilon^2\right)$.
\end{theorem}

\begin{proof}

When the dimension $d \leq \mathcal{O}( \log^2 n)$, we skip the FJLT and directly plug in $m=d$ to Lemma \ref{lem:4}. This tells us that the Quadtree step of the algorithm outputs estimators with expected distortion $F = \OO(\min(d, \sqrt{d}\log n)) = \OO(\log^2 n)$. Applying Theorem \ref{thm:adapted} on this $F$, we get a total runtime of $\OO(nd(\log\log n + \log (1/\varepsilon))/\varepsilon^2)$ as claimed.

On the other hand, in the high-dimension regime ($d \geq \Omega( \log^2 n)$), we choose appropriate $m = \OO(\log^2n)$ and runs the FJLT from Lemma \ref{lem:reduction}. This takes $\OO(nd \log\log n)$ time. Then we are back in the low-dimensional case, where we spend an additive $\OO(nd(\log\log n + \log (1/\varepsilon))/\varepsilon^2)$ time to complete running the algorithm. 

\end{proof}

\section{Extensions and Theoretical Limits}\label{sec:extension}

To substantiate the broad applicability of our techniques and to map the theoretical boundaries of the approximation, we introduce the application to non-geometric norms (LLM Retrieval metrics) and prove conditional lower bounds for the memory required by streaming models. Due to the page limit, we delay the proofs to the appendix \cref{sec:appendix-extension}.

\subsection{MaxSim and the ColBERT}

In Natural Language Processing, the ColBERT model utilizes a similarity measure intrinsically related to the Chamfer distance. 

\begin{definition}[ColBERT MaxSim]\label{def:colbert}
For embedded document tokens $A, B \subset \mathbb{S}^{d-1}$ on the unit hypersphere, the ColBERT score is defined as:
\begin{equation}
\text{MaxSim}(A,B) = \sum_{a \in A} \max_{b \in B} \langle a, b \rangle.
\end{equation}
\end{definition}

\begin{theorem}[Fast Approximate ColBERT Retrieval]\label{thm:colbert}
For normalized sets $A, B \subset \mathbb{S}^{d-1}$, the Fast $\ell_2$ Chamfer distance algorithm (Theorem \ref{thm:l2_chamfer}) can be modified to output an additive $\mathcal{O}(\varepsilon n)$-approximation to $\text{MaxSim}(A,B)$, and for $d \geq \poly\log (n)$, it runs in $\mathcal{O}\left(nd(\log \log n + \log(1/\varepsilon))/\varepsilon^2\right)$ time.
\end{theorem}



\subsection{Streaming Algorithm and Space Complexity Lower Bound}\label{sec:streaming}

In large-scale data settings, the input sets $A$ and $B$ are often not static but arrive as a data stream with coordinate updates. We analyze the Chamfer distance under the dynamic turnstile stream model, where points receive additive updates to their coordinates. 

\textbf{Upper Bound via Linear Sketching:} Our dimensionality reduction techniques are naturally suited for this environment. Since the projection matrices $T$ used in our algorithms (both Toeplitz FJLT and Lopsided embeddings) are linear, data-oblivious maps, they trivially support additive updates: $T(v + \Delta) = Tv + T\Delta$. This allows the algorithm to maintain the low-dimensional embeddings online. Once the stream terminates, the solver of \cite{Feng2025} is applied black-box to the sketches. This dynamic algorithm reduces the memory footprint from $\mathcal{O}(nd)$ raw coordinates to $\mathcal{O}(nm) = \mathcal{O}(n \text{ poly}(\log n))$ space.

We formalize that this $\tilde{\mathcal{O}}(n)$ space complexity is near-optimal by proving an unconditional $\Omega(n)$ space lower bound via communication complexity. 

\begin{theorem}[$\Omega(n)$ Space Lower Bound for Streaming Chamfer]\label{thm:lower}
Any randomized constant-pass streaming algorithm that computes a constant-factor approximation to the $\ell_p$ Chamfer distance ($p \ge 1$) for dynamically updated point sets requires $\Omega(n)$ bits of space.
\end{theorem}

\section{Experiments} \label{sec:experiments}

We evaluate the performance of our algorithm on both synthetic high-dimensional data and real-world natural language embeddings. We focus on three dimensions of performance: time complexity, space complexity, and approximation accuracy (validating the combination of lopsided embeddings / FJLT with the Chamfer algorithm. \footnote{In our implementations, instead of using \cite{Feng2025} as the black-box Chamfer algorithm as in our theoretical proofs, we use \cite{BIJ24}, because efficient implementation of \cite{Feng2025} requires optimizing the efficiency of bit-level operations, which is beyond the focus of our experiments. This also demonstrates that our embeddings and dimension reduction approach accelerates any downstream $\ell_1$ solver.}). 

\subsection{Experimental Setup}
\textbf{Baselines:} We compare our method against two baselines. First, the \textbf{Naive Exact} method, which computes the full $\mathcal{O}(n^2 d)$ distance matrix. Second, the unsketched Chamfer algorithm from  \cite{BIJ24}, which runs QuadTrees and uses the outputted estimator to perform importance sampling in the original high-dimensional space without our random projections, theoretically taking $\mathcal{O}(nd \log n)$ time.

\textbf{Datasets:} For computational scaling experiments, we generate synthetic sets $A, B \subset \mathbb{R}^d$ with $d=2048$ to accentuate the high-dimensional bottlenecks. For accuracy validation, we use the 300-dimensional GloVe dataset \cite{Kusner2015} consisting of word embeddings, a standard benchmark for NLP metrics like the ColBERT retrieval score (Definition \ref{def:colbert}).
All experiments were conducted on a single NVIDIA T4 GPU. Our algorithm projects to a target dimension of $m=64$ and uses $s \approx \mathcal{O}(1/\varepsilon^2)$ importance samples for the importance sampling step of   \cite{BIJ24}. 

\begin{figure*}[t]
\centering
\begin{minipage}{0.62\textwidth}
  \includegraphics[width=\linewidth]{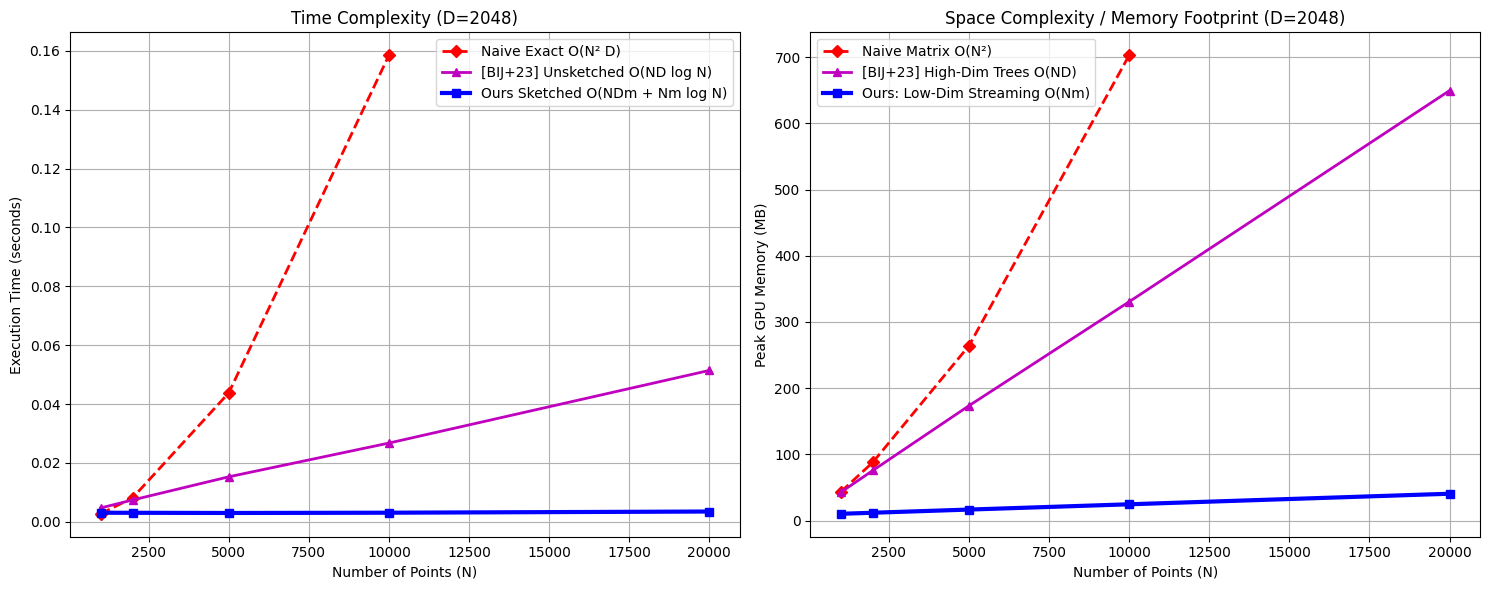}
\end{minipage}
\begin{minipage}{0.36\textwidth}
  \includegraphics[width=\linewidth]{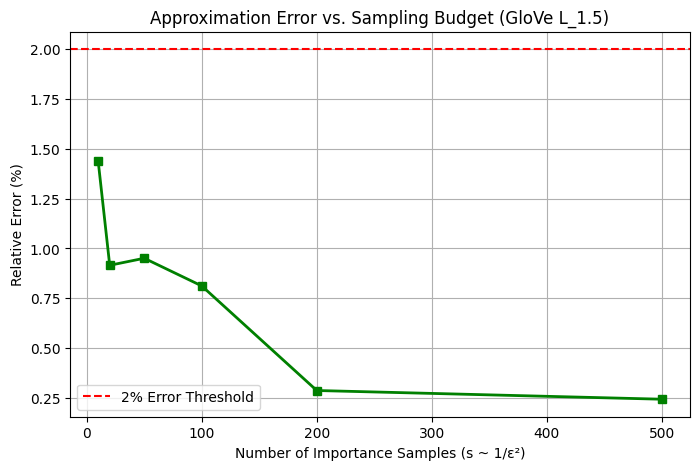}
\end{minipage}
\caption{\textbf{Left:} Runtime scaling ($D=2048$). Our method remains flat near the GPU noise-floor, i.e., the execution time is consistently $0.005$ across different number of points. \textbf{Middle:} Peak GPU memory ($D=2048$). Our method validates the $\tilde{\mathcal{O}}(nm)$ space claim, using less than 45 MB while previous methods exceed 600 MB. \textbf{Right:} Approximation error on the GloVe dataset for the $\ell_{1.5}$ Chamfer distance, showing sub-1\% error with very few importance samples.}
\label{fig:results}
\end{figure*}

\subsection{Computational Efficiency (Time Scaling)}
We evaluate the wall-clock execution time as $n$ increases from $1,000$ to $20,000$ with a fixed dimension $d=2048$. As shown in Figure \ref{fig:results} (Left), the exact method scales quadratically, running out of memory  beyond $n=10,000$. 

The unsketched  \cite{BIJ24} baseline scales linearly but incurs a high constant overhead due to operating directly in $\mathbb{R}^{2048}$, taking 51ms at $n=20,000$. In contrast, our algorithm runs in $5$ms for $n=20,000$. 


\subsection{Space Complexity and Streaming Feasibility}
In Section \ref{sec:streaming}, we claimed our algorithm supports dynamic streams via a low-memory footprint of $\tilde{\mathcal{O}}(nm)$. Figure \ref{fig:results} (Middle) validates this claim by tracking Peak GPU Memory allocation.

The exact distance matrix requires $\mathcal{O}(n^2)$ space, immediately exhausting memory. Also, \cite{BIJ24} must retain the high-dimensional vectors $A$ and $B$ to construct the quadtrees, consuming 649 MB at $n=20,000$. Our method operates strictly on the sketches. By discarding the high-dimensional inputs after the oblivious random projection, our peak memory consumption is just 41 MB at $n=20,000$, which is roughly $ 15\times$ reduction over the linear unsketched baseline. This empirically confirms the efficiency of our algorithm in memory-constrained environments.

\subsection{Accuracy and Lopsided Embedding Validation}
In \cref{tab:accuracy}, we record the relative approximation error of our algorithm against the unsketched baseline \cite{BIJ24}. As expected, the unsketched baseline achieves smaller error because it operates directly in the ambient space $\R^{2048}$, whereas our method introduces distortion through dimensionality reduction. At the same time, the distortion remains small in practice: in these experiments, the relative error of our method stays below 1\%, with the largest mean error peaking at just 0.622\%.

\begin{table*}[t]
\centering
\caption{Mean relative error (\%) for \cite{BIJ24} and our method across different values of $N$. Results are reported as mean $\pm$ standard deviation.}
\label{tab:accuracy}
\resizebox{0.68\textwidth}{!}{%
\begin{tabular}{@{}ccc@{}}
\toprule
\textbf{$N$} & \textbf{\cite{BIJ24} Mean Relative Error (\%)} & \textbf{Ours Mean Relative Error (\%)} \\
\midrule
1000  & 0.055 $\pm$ 0.023 & 0.614 $\pm$ 0.412 \\
2000  & 0.066 $\pm$ 0.062 & 0.559 $\pm$ 0.058 \\
5000  & 0.067 $\pm$ 0.040 & 0.622 $\pm$ 0.249 \\
10000 & 0.027 $\pm$ 0.025 & 0.397 $\pm$ 0.264 \\
\bottomrule
\end{tabular}%
}
\end{table*}

We also validate the accuracy of combining lopsided $\ell_p \to \ell_1$ embeddings with the Chamfer algorithm (Theorem \ref{thm:general_case}, we compute the $\ell_{1.5}$ Chamfer distance on GloVe word embeddings. 

Figure \ref{fig:results} (Right) plots the relative approximation error against the importance sampling budget $s$. Following our theoretical guarantees, the error decays rapidly. With a sample budget of just $s=10$, the error is $1.44\%$. Increasing the budget to $s=100$ drops the error to $0.81\%$, and at $s=500$, the error reaches $0.24\%$. 

This confirms one of the theoretical novelties of our work, that the weak guarantees of the lopsided embeddings (non-contraction and bounded first moment, Lemma \ref{lemma:lopsided}) are sufficient to preserve the Chamfer distance from $\ell_p$ to $\ell_1$.

\section{Conclusion}
We give efficient algorithms for approximating the Chamfer distance. For $1 < p < 2$, we use lopsided $\ell_p$ to $\ell_1$ embeddings and show that they reduce the $\ell_p$ Chamfer distance instance to $\ell_1$ with small distortion. This lead to $\mathcal{O}(nd\cdot \poly(\log \log n, \log(1/\varepsilon))/\varepsilon^2)$ time algorithm for estimating the Chamfer distance. 
For $p = 2$, we successfully resolved the open problem by employing the FJLT using Toeplitz matrices \cite{Krahmer2011} and integrating it with the framework of \cite{Feng2025}. This yields a runtime of $\mathcal{O}(nd(\log \log n + \log(1/\varepsilon))/\varepsilon^2)$, matching the state-of-the-art $\ell_1$ result.

We also connect our fast $\ell_2$ Chamfer algorithms to MaxSim-style LLM retrieval (e.g., ColBERT), obtaining a near-linear-time
additive approximation. Finally, in the dynamic turnstile model, the
linearity of our embeddings gives $\tilde{\OO}(n)$-space streaming algorithms, and we prove a matching $\Omega(n)$ space lower bound
for any randomized constant-pass constant-factor approximation (for all $p\ge 1$).

Regarding the limitations of our work, in the plain model, our result leaves a factor of $\log\log n$ between our upper bound and the na\"ive $\Omega(nd)$ lower bound. It would be interesting to remove this overhead, or else prove a matching lower bound. Empirically, our evaluation is promising but limited: it uses synthetic data plus GloVe, a single GPU setup. It would be interesting to evaluate our algorithm on a broader range of large-scale point-cloud and retrieval datasets.
\paragraph{Impact Statement} This paper presents work whose goal is to advance the field of Machine
Learning. There are many potential societal consequences of our work, none
which we feel must be specifically highlighted here.
 
\paragraph{Acknowledgements}
Ying Feng was supported by the NSF TRIPODS program (award DMS-2022448).  David P. Woodruff would like to acknowledge support from a Simons Investigator Award and Office of Naval Research (ONR) award number N000142112647.
\paragraph{AI Acknowledgements}
We used Gemini to assist with drafting, revising, and checking the exposition, mathematical arguments, and experimental code. The tools materially affected all sections of the paper. The starting point of this work was a query to an internal version of Gemini asking it to extend the algorithm in \cite{Feng2025} from $\ell_1$ to $\ell_2$. By combining the model's output with an existing fast JL transform, we obtained the result for the $\ell_2$ case. For the remaining results, the human authors provided the initial outlines, while the proofs were derived by Gemini. The authors verified the correctness and originality of all content, including references.

\bibliographystyle{alpha}
\bibliography{references}

\newcommand{\etalchar}[1]{$^{#1}$}
\begin{thebibliography}{AMN{\etalchar{+}}98}

\bibitem[AM19]{atasu19a}
Kubilay Atasu and Thomas Mittelholzer.
\newblock Linear-complexity data-parallel earth mover's distance approximations.
\newblock In Kamalika Chaudhuri and Ruslan Salakhutdinov, editors, {\em Proceedings of the 36th International Conference on Machine Learning}, volume~97 of {\em Proceedings of Machine Learning Research}, pages 364--373. PMLR, 09--15 Jun 2019.

\bibitem[AMN{\etalchar{+}}98]{AM98}
Sunil Arya, David~M. Mount, Nathan~S. Netanyahu, Ruth Silverman, and Angela~Y. Wu.
\newblock An optimal algorithm for approximate nearest neighbor searching fixed dimensions.
\newblock {\em J. ACM}, 45(6):891–923, November 1998.

\bibitem[AR15]{andoni2015optimal}
Alexandr Andoni and Ilya Razenshteyn.
\newblock Optimal data-dependent hashing for approximate near neighbors.
\newblock In {\em Proceedings of the forty-seventh annual ACM symposium on Theory of computing}, pages 793--801, 2015.

\bibitem[AS03]{athitsos2003estimating}
Vassilis Athitsos and Stan Sclaroff.
\newblock Estimating 3d hand pose from a cluttered image.
\newblock In {\em 2003 IEEE Computer Society Conference on Computer Vision and Pattern Recognition, 2003. Proceedings.}, volume~2, pages II--432. IEEE, 2003.

\bibitem[BIJ{\etalchar{+}}23]{BIJ24}
Ainesh Bakshi, Piotr Indyk, Rajesh Jayaram, Sandeep Silwal, and Erik Waingarten.
\newblock A near-linear time algorithm for the chamfer distance, 2023.

\bibitem[EK23]{elkin23a}
Yury Elkin and Vitaliy Kurlin.
\newblock A new near-linear time algorithm for k-nearest neighbor search using a compressed cover tree.
\newblock In Andreas Krause, Emma Brunskill, Kyunghyun Cho, Barbara Engelhardt, Sivan Sabato, and Jonathan Scarlett, editors, {\em Proceedings of the 40th International Conference on Machine Learning}, volume 202 of {\em Proceedings of Machine Learning Research}, pages 9267--9311. PMLR, 23--29 Jul 2023.

\bibitem[FI25]{Feng2025}
Ying Feng and Piotr Indyk.
\newblock Even faster algorithm for the chamfer distance.
\newblock In {\em International Colloquium on Automata, Languages and Programming}, 2025.

\bibitem[FSG17a]{Fan2017}
Haoqiang Fan, Hao Su, and Leonidas~J Guibas.
\newblock A point set generation network for 3d object reconstruction from a single image.
\newblock In {\em Proceedings of the IEEE conference on computer vision and pattern recognition}, pages 605--613, 2017.

\bibitem[FSG17b]{fan2017point}
Haoqiang Fan, Hao Su, and Leonidas~J Guibas.
\newblock A point set generation network for 3d object reconstruction from a single image.
\newblock In {\em Proceedings of the IEEE conference on computer vision and pattern recognition}, pages 605--613, 2017.

\bibitem[GJK{\etalchar{+}}25]{G25}
Gramoz Goranci, Shaofeng H-C Jiang, Peter Kiss, Eva Szilagyi, and Qiaoyuan Yang.
\newblock Fully dynamic algorithms for chamfer distance.
\newblock In {\em The 39th Annual Conference on Neural Information Processing Systems}, 2025.

\bibitem[HV11]{Hinrichs2011}
Aicke Hinrichs and Jan Vyb{\'\i}ral.
\newblock Johnson--lindenstrauss lemma for circulant matrices.
\newblock {\em Random Structures \& Algorithms}, 39(3):391--398, 2011.

\bibitem[JL84]{Johnson1984}
William~B Johnson and Joram Lindenstrauss.
\newblock Extensions of lipschitz mappings into a hilbert space.
\newblock {\em Contemporary mathematics}, 26(189-206), 1984.

\bibitem[JSQJ18]{jiang2018gal}
Li~Jiang, Shaoshuai Shi, Xiaojuan Qi, and Jiaya Jia.
\newblock Gal: Geometric adversarial loss for single-view 3d-object reconstruction.
\newblock In {\em Proceedings of the European conference on computer vision (ECCV)}, pages 802--816, 2018.

\bibitem[KSKW15]{Kusner2015}
Matt Kusner, Yu~Sun, Nicholas Kolkin, and Kilian Weinberger.
\newblock From word embeddings to document distances.
\newblock In {\em International conference on machine learning}, pages 957--966. PMLR, 2015.

\bibitem[KW11]{Krahmer2011}
Felix Krahmer and Rachel Ward.
\newblock New and improved johnson--lindenstrauss embeddings via the restricted isometry property.
\newblock {\em Journal on Mathematical Analysis}, 43(3):1269--1281, 2011.

\bibitem[KZ20]{khattab2020colbert}
Omar Khattab and Matei Zaharia.
\newblock Colbert: Efficient and effective passage search via contextualized late interaction over bert.
\newblock In {\em Proceedings of the 43rd International ACM SIGIR conference on research and development in Information Retrieval}, pages 39--48, 2020.

\bibitem[LLW23]{Li2023}
Yi~Li, Honghao Lin, and David~P. Woodruff.
\newblock $\ell_p$-regression in the arbitrary partition model of communication, 2023.

\bibitem[LRSS25]{lahn25a}
Nathaniel Lahn, Sharath Raghvendra, Emma Saarinen, and Pouyan Shirzadian.
\newblock Scalable approximation algorithms for $p$-{W}asserstein distance and its variants.
\newblock In Aarti Singh, Maryam Fazel, Daniel Hsu, Simon Lacoste-Julien, Felix Berkenkamp, Tegan Maharaj, Kiri Wagstaff, and Jerry Zhu, editors, {\em Proceedings of the 42nd International Conference on Machine Learning}, volume 267 of {\em Proceedings of Machine Learning Research}, pages 32188--32205. PMLR, 13--19 Jul 2025.

\bibitem[SMFW04]{sudderth2004visual}
Erik~B Sudderth, Michael~I Mandel, William~T Freeman, and Alan~S Willsky.
\newblock Visual hand tracking using nonparametric belief propagation.
\newblock In {\em 2004 Conference on Computer Vision and Pattern Recognition Workshop}, pages 189--189. IEEE, 2004.

\bibitem[SW26]{Swartworth2026}
William Swartworth and David~P Woodruff.
\newblock Sketching faster than dimension times update time.
\newblock {\em Manuscript}, 2026.

\bibitem[Vyb11]{Vybiral2011}
Jan Vyb{\'\i}ral.
\newblock A variant of the johnson--lindenstrauss lemma for circulant matrices.
\newblock {\em Journal of Functional Analysis}, 260(4):1096--1105, 2011.

\end{thebibliography}

\newpage
\appendix

\section{Proof to Lemma \ref{lemma:lopsided}}\label{sec:appendix}

\begin{lemma}[Repeating Lemma \ref{lemma:lopsided}] Suppose $p \in (1, 2)$ is constant. Let $\varepsilon, \delta \in (0, 1/2)$ and $V \subset \mathbb{R}^d$ with $|V| = n$. There exists a distribution over matrices $T \in \mathbb{R}^{m \times d}$, where $m = \poly(\log n, \varepsilon^{-1}, \delta^{-1})$, satisfying:
\begin{enumerate}
    \item \textbf{(Expected Error Bound):} For any fixed $v \in \mathbb{R}^d$, $\mathbb{E}[|\|Tv\|_1 - \|v\|_p|] \le \varepsilon\|v\|_p$.
    \item \textbf{(Contraction on V):} With probability $1 - \delta$, for all $v \in V$, $\|Tv\|_1 \ge (1 - \varepsilon)\|v\|_p$.
\end{enumerate}
\end{lemma}

\begin{proof}
    In  Lemma 18 of \cite{Li2023}, this  is proven for $\|Tv\|_1$ replaced with $\|Tv\|_q$ for any $q \in (1, p)$. We trace the proof of Lemma 18 and verify that setting $q=1$ does not violates any step of the proof.

\textbf{Step 1. Moment Existence.}
We define $T$ to be a matrix whose entries  are i.i.d. $p$-stable random variables. It is a fundamental property of $p$-stable distributions that the $q$-th moment $\mathbb{E}[|X|^q]$ is finite if and only if $q < p$. This implies Lemma 25 of \cite{Li2023} for $q = 1$, that 
\[\E[\|Tv\|_1] = \E[\sum_{i =1}^m |(Tv)_i|] =  m \alpha_{p} \|v\|_p \]
for some constant $\alpha_p$ depending only on $p$.  Scaling $T$ by $1/(m \cdot \alpha_{p})$ gives $\E[\sum_{i =1}^m |(Tv)_i|] =  \|v\|_p $.    Similarly, it gives $\E[| (Tv)_i|^r] =\beta_{p, r}  \|v\|^r_p/m^r $ for every $i$ for some constant $\beta_{p, r}$ depending only on $p, r$.

\textbf{Step 2. Concentration via von Bahr-Esseen.}
To show concentration, the proof selects a higher moment $r \in (1, p)$.
The proof then considers the random variables $X_i = |(Tv)_i| - \mathbb{E}[|(Tv)_i|]$, which are independent and mean-zero. 
We apply the von Bahr-Esseen inequality (Lemma 27 of \cite{Li2023}), which bounds $\mathbb{E}[|\sum X_i|^s]$ for $1 \le s \le 2$. Here, the exponent is chosen to be $s = r/1 = r \le 2$. The  von Bahr-Esseen inequality  applies perfectly and provides:
$$ \mathbb{E}\left[\left||\sum_{i =1}^m (Tv)_i| - \mathbb{E}[|\sum_{i =1}^m (Tv)_i|]\right|^{r}\right] \le 2 \sum_{i=1}^m \mathbb{E}\left[ \left| |(Tv)_i| - \frac{1}{m}\|v\|_p \right|^{r} \right] $$

\textbf{Step 3. Bounding the Central Moments.}
Proposition 26 of \cite{Li2023} gives 
\[\mathbb{E} ||X| - \mathbb{E}|X||^r \le 2^r \mathbb{E}|X|^{r}.\]
Applying this with $X = (Tv)_i$, the RHS is $2^{r} \mathbb{E}[|(Tv)_i|^{r}] = \OO(m^{-r} \|v\|_p^{r})$. Summing over the $m$ terms gives:
$$ \mathbb{E}[|\|Tv\|_1 - \mathbb{E}[\|Tv\|_1]|^{r}] \le \OO(m^{1-r} \|v\|_p^{r}).$$ Then  using standard moment inequalities, we can bound the first moment by higher moments: for any $r > 1$, \[\mathbb{E}[|S - \mathbb{E}[S]|] \le (\mathbb{E}[|S - \mathbb{E}[S]|^r])^{1/r}.\] Plugging in $S = \|Tv\|_1$ and $m \geq \poly(\varepsilon^{-1})$, we see that the error $\mathbb{E}[|\|Tv\|_1 - \|v\|_p|] $ is bounded by $\mathcal{O}(\varepsilon \|v\|_p)$ as desired.
This concludes the proof to property (1).

Moving to property (2), we note that this part of the original proof to Lemma 18 is independent of the choice of $q$. Therefore, the only difference is that: property (2) of Lemma 18 in \cite{Li2023} argues for an entire subspace by union bounding over a $\gamma$-net, while in our case this is simplified to a union bound over a finite set $V$. This is supported as long as $m$ is chosen to be appropriate $\poly(\log n, \eps^{-1}, \delta^{-1})$.

\end{proof}

\section{Delayed Proofs from \cref{sec:quadtree}}\label{sec:appendix-quadtree}

In this section, we repeat the statements of the lemmas and give the proofs.

\begin{lemma}[Repeating Lemma \ref{lem:1}]

For all $a, b \in \R^m$ and $0 \leq k \leq t$, if $h_k(a) = h_k(b)$, then $\|a -b\|_2 \leq 2^k \sqrt{m}$.
\end{lemma}

\begin{proof}
    If $h_k(a) = h_k(b)$, then $a, b$ lie in the same $m$-dimensional cube of side-length $2^k$. The $\ell_2$ diameter of such a cube is $2^k \sqrt{m}$.
\end{proof}
\begin{lemma}[Repeating Lemma \ref{lem:2}]
    With probability $1 - \OO(1/n)$, the following holds
simultaneously for all $a,b, k$: If $h_k(a) = h_k(b)$, then $\|a-b\|_2 \leq 2^k \cdot 3\log n$.
\end{lemma}
\begin{proof}
    Lemma 3.4 of \cite{Feng2025} shows that $\forall a,b, k: \|a-b\|_1 \leq 2^k \cdot 3\log n$ with probability  $1 - \OO(1/n)$. Combining this with the fact that $\|a-b\|_1 \geq \|a-b\|_2$ concludes the proof.
\end{proof}

\begin{lemma}[Repeating Lemma \ref{lem:4}]
    With probability $1 - \OO(1/n)$, it holds for all $a \in A$ that $\E{\mathcal{D}_a} \leq F \cdot \mathsf{opt}(a)$, where $F  = \OO(\min(m,\sqrt{m}\log n))$.
\end{lemma}

\begin{proof}
 We need the following fact:
\begin{fact}[\cite{BIJ24}]
     $\Pr[h_k(a) \neq h_k(b)] \leq \frac{\|a-b\|_1}{2^k}.$
\end{fact}
    Similar to Theorem 3.5 in \cite{Feng2025}, we condition on Lemma \ref{lem:2} and fix $a \in A$.  We also fix $b^* := \arg\min_{b\in B} \|a-b\|_2$ and define $k^* := \lceil \log (\|a-b^*\|_1) \rceil$. 
    
    Let $\mathcal{E}_k$ be the event that $\tilde{k} = k$ (recall that we identified a unique $\tilde{k}$ when defining the estimator). If $\mathcal{E}_k$  occurs, the estimator $\mathcal{D}_a$ is bounded by $2^k \cdot \tilde{F}$ for $\tilde{F} = \OO(\min(\sqrt{m}, \log n))$. Thus 
    \[\E{\mathcal{D}_a} \leq \sum_{0 \leq k \leq t}\Pr[\mathcal{E}_k]\cdot 2^k \cdot \tilde{F} .\] We split the sum based on $k \leq k^*$ and $k > k^*$. The the first part by 
    \[\tilde{F} 2^{k^*} \sum_{k \leq k^*} \Pr[\mathcal{E}_k] \leq \tilde{F} \|a-b^*\|_1.\] For the second part, 
    \begin{align*}
        \sum_{k > k^*}\Pr[\mathcal{E}_k]\cdot 2^k \cdot \tilde{F}  &\leq \tilde{F}  \sum_{k > k^*}(\frac{\|a-b^*\|_1}{2^{k-1}})^2 \cdot 2^k\\ 
        &\leq 4\tilde{F}\cdot \|a-b^*\|_1  \sum_{k > k^*}\frac{\|a-b^*\|_1}{2^k} \\
        &= \OO(\tilde{F}\|a-b^*\|_1).
    \end{align*}
    Thus, $\E{\mathcal{D}_a} = \OO(\tilde{F})\cdot \|a-b^*\|_1 \leq \OO(\tilde{F}\sqrt{m})\cdot \mathsf{opt}_a$. The approximation factor is $F = \OO(\min(m,\sqrt{m}\log n))$.
    
\end{proof}

\section{Delayed Proofs from  \cref{sec:extension}}\label{sec:appendix-extension}

In this section, we repeat the statements of the theorems and give the proofs.

\begin{theorem}[Repeating \cref{thm:colbert}]
For normalized sets $A, B \subset \mathbb{S}^{d-1}$, the Fast $\ell_2$ Chamfer distance algorithm (Theorem \ref{thm:l2_chamfer}) can be modified to output an additive $\mathcal{O}(\varepsilon n)$-approximation to $\text{MaxSim}(A,B)$, and for $d \geq \poly\log (n)$, it runs in $\mathcal{O}\left(nd(\log \log n + \log(1/\varepsilon))/\varepsilon^2\right)$ time.
\end{theorem}
\begin{proof}
For any $a, b \in \mathbb{S}^{d-1}$, the geometric identity $\|a-b\|_2^2 = \|a\|_2^2 + \|b\|_2^2 - 2\langle a, b \rangle = 2 - 2\langle a, b \rangle$ holds. Maximizing the inner product is strictly equivalent to minimizing the squared Euclidean distance. Therefore, we can output an approximation $|A| -  \frac{1}{2}\tilde{\CH}_2(A, B)$, where $\tilde{\CH}_2(A, B)$ is an $(1+\varepsilon)$-approximation to $\sum_{a \in A} \min_{b \in B }\|a-b\|^2_2$. Note that this quantity corresponds to the Chamfer distance with the underlying metric being the squared $\ell_2$ norms (as opposed to $\ell_2$).

To approximate this, we revisit Lemma 4.3, the analysis of
the approximation factor $F$ for the QuadTree estimator $D_a$.
The modification is to use three independently shifted
quadtrees instead of two, and to let $D_a$ denote the squared
Euclidean distance to the returned candidate.

Fix $a\in A$, let $b^*=\arg\min_{b\in B}\|a-b\|_2$,
and define $\Delta=\|a-b^*\|_1$ and
$k^*=\lceil\log_2\Delta\rceil$.
Let $E_k$ denote the event that the smallest collision level
among the three trees is $k$.
On $E_k$, we have $D_a\le m2^{2k}$.

For $k>k^*$, the event $E_k$ implies that $a$ and $b^*$
are separated at level $k-1$ in all three trees.
By Fact C.4 and independence,
\[
\Pr[E_k]\le
\left(\frac{\Delta}{2^{k-1}}\right)^3.
\]
Thus,
\[
\begin{aligned}
\mathbb E[D_a]
&\le \sum_{k\le k^*}m2^{2k}
 +\sum_{k>k^*}\Pr[E_k]m2^{2k}\\
&\le O(m\Delta^2)
 +8m\Delta^3\sum_{k>k^*}2^{-k}\\
&=O(m\Delta^2)
 =O(m^2\|a-b^*\|_2^2),
\end{aligned}
\]
where the last step uses
$\|a-b^*\|_1\le\sqrt m\|a-b^*\|_2$.

Therefore, the expected approximation factor for
squared Euclidean distances is $F=O(m^2)$.
The remaining rejection- and importance-sampling analysis
of [FI25] applies with only constant-factor changes:
squaring a factor-$2$ approximate nearest-neighbor
distance gives a factor-$4$ estimate.
The rest of the algorithm from Theorem 4.4,
including the FJLT, remains unchanged.
Since $m=O(\log^2 n)$, the total runtime is
$O(nd(\log\log n+\log(1/\varepsilon))/\varepsilon^2)$
for $d\ge\operatorname{poly}\log(n)$, as claimed.

Finally, the relative approximation error for the
squared Euclidean Chamfer distance translates to
an additive $O(\varepsilon n)$ approximation to MaxSim,
following the geometric identity and the unit sphere
assumption.

\end{proof}

\begin{theorem}[Reapeating \cref{thm:lower}]
Any randomized constant-pass streaming algorithm that computes a constant-factor approximation to the $\ell_p$ Chamfer distance ($p \ge 1$) for dynamically updated point sets requires $\Omega(n)$ bits of space.
\end{theorem}

\begin{proof}
We reduce from the 2-Player Set Disjointness (DISJ) problem, which is a core hardness result in communication complexity. Alice and Bob receive indicator vectors $x, y \in \{0,1\}^n$ respectively. We assume the standard promise that the Hamming weights are $|x|_1 = |y|_1 = n/4$. DISJ asks whether $x \cap y = \emptyset$ (i.e., whether there exists some index $i$ such that $x_i = y_i = 1$). It is a well-known result that solving DISJ requires $\Omega(n)$ communication.

We map this to a dynamic point stream in $\mathbb{R}^2$. Let the target set be a fixed multiset $B$ containing $ n/4$ copies of $(1,0)$,   $ n/4$ copies of $(0,1)$, and  $ n/2$ copies of $(0,0)$. Note that $|B| = n$.
Let the source set $A = \{a_1, \dots a_n\}$ be initialized with $n$ points at the origin $(0,0)$. 

In the stream, Alice receives $x$ and applies additive updates to the first coordinate of the $i$-th point of $A$, setting $a_i = (x_i, 0)$. She then transmits the memory state of the streaming data structure to Bob. Bob receives $y$ and applies additive updates to the second coordinate, resulting in the final point set $A = \{(x_i, y_i) \mid i \in [n]\}$. 

We analyze the $\ell_p$ Chamfer distance $\text{CH}_p(A,B)$:
\begin{itemize}
    \item \textbf{If the sets are Disjoint ($x \cap y = \emptyset$):} The condition $x_i \cdot y_i = 0$ holds for all $i$. Due to the fixed Hamming weights, the resulting set $A$ contains exactly $n/4$ points at $(1,0)$, $n/4$ points at $(0,1)$, and $n/2$ points at $(0,0)$. Thus, $A$ and $B$ are identical multisets, and the distance $\text{CH}_p(A,B) = 0$.
    \item \textbf{If the sets Intersect ($x \cap y \neq \emptyset$):} There exists at least one index $i$ where $(x_i, y_i) = (1,1)$. The $\ell_p$ distance from the point $(1,1)$ to any point in the reference set $B$ is $\min(\|(1,1)-(0,0)\|_p, \|(1,1)-(1,0)\|_p, \|(1,1)-(0,1)\|_p) = \min(2^{1/p}, 1, 1) = 1$. Thus, $\text{CH}_p(A,B) \ge 1$.
\end{itemize}

Since distinguishing between a distance of $0$ and a distance $\ge 1$ allows one to decide the Set Disjointness problem, the space required to store the streaming data structure (transmitted from Alice to Bob) must be at least the communication complexity of DISJ. Thus, the space required is $\Omega(n)$ bits.

\end{proof}


\end{document}